\documentclass[aps,superscriptaddress,twocolumn,nofootinbib]{revtex4}

\usepackage{graphics}
\input{epsf}

\begin{document}

\title{Inflation in the warm and cold regimes}

\author{Arjun Berera}      
\email{ab@ph.ed.ac.uk}
\affiliation{School of Physics, University of Edinburgh, 
Edinburgh EH9 3JZ, United Kingdom}

\begin{abstract}
It is now understood that inflation dynamics comes
in two forms, isentropic or cold inflation and nonisentropic or warm
inflation. In the former, inflation occurs without radiation production,
whereas in the latter both radiation production and
inflation occur concurrently.  Recent, detailed, quantum field
theory calculations have shown that many generic inflation
models, including hybrid inflation, which were believed only
to have cold inflation regimes, in fact have regimes of both warm
and cold inflation.  These results dispel many foregone assumptions
generally made up to now about inflation models and bring to
the fore various elementary issues that must be addressed to
do reliable calculations from inflation models.
Here I review these results and issues.
I then show that warm inflation has intrinsic model independent features
that makes it natural or equivalently have no ``eta problem''.
Next density perturbations and observational consequences of warm
inflation are discussed.
Finally the implications of warm inflation to model building
and physics beyond the Standard Model are outlined.

\medskip

\medskip

Plenary talk at Cosmion04. appears in conference proceedings,

Gravitation and Cosmology, V. 11, No. 1-2 (41-42), 51 (2005).

\end{abstract}

\maketitle




\section{Introduction}
\label{sect1}

For over two decades, inflation has been a very successful idea.
In its earliest days, this success was attributed to the ability for
this idea to unite particle physics and cosmology.  In the
past decade, the success of the inflation idea has been driven
by its consistency to observation based on data in particular
from precision CMB satellite experiments.
The growing success over the years of the inflation idea has also led
to an increasing understanding of the underlying dynamics
of inflation.  By now it is known that there are two
dynamical realizations of inflation.  One is the
original or standard picture, also referred to as isentropic
or cold inflation \cite{oldi,ni,ci}.
In this picture inflationary expansion occurs with the universe
in a supercooled phase which subsequently ends with a reheating
period that introduces radiation into the universe.
The fluctuations created during inflation are effectively
zero-point ground state fluctuations and the evolution of the inflaton
field is governed by a ground state evolution equation.
In an alternative class of cold inflation models,
inflation has a geometrical origin \cite{star}
with adiabatic density perturbations \cite{mc}.
The other picture of inflation dynamics is nonisentropic
or warm inflation \cite{wi}.  In this picture, 
inflationary expansion and radiation
production occur concurrently.  Moreover, the fluctuations created
during inflation emerge from some excited statistical state and the
evolution of the inflaton has dissipative terms arising from the
interaction of the inflaton with other fields.

The dividing point between warm and cold inflation is roughly at
$\rho_r^{1/4} \approx H$, where $\rho_r$ is the radiation energy density
present during inflation and $H$ is the Hubble parameter. Thus
$\rho_r^{1/4} > H$ is the warm inflation regime and $\rho_r^{1/4} < H$
is the cold inflation regime. This criteria is independent of
thermalization, but if such were to occur, one sees this criteria
basically amounts to the warm inflation regime corresponding to when $T
> H$. This is easy to understand since the typical inflaton mass during
inflation is $m_\phi \approx H$ and so when $T>H$, thermal fluctuations
of the inflaton field
will become important. This criteria for entering the warm inflation
regime turns out to require the dissipation of a very tiny fraction of
the inflaton vacuum energy during inflation. {}For example, for
inflation with vacuum ({\it i.e.} potential) energy at the GUT scale
$\sim 10^{15-16} {\rm GeV}$, in order to produce radiation at the scale
of the Hubble parameter, which is $\approx 10^{10-11} {\rm GeV}$, it just
requires dissipating one part in $10^{20}$ of this vacuum energy density
into radiation. Thus energetically not a very significant amount of
radiation production is required to move into the warm inflation regime.
In fact the levels are so small, and their eventual effects
on density perturbations and inflaton evolution are so significant, that
care must be taken to account for these effects in the analysis of any
inflation models.

In this paper, we will examine warm inflation dynamics
and compare it to cold inflation dynamics.
In Sect. \ref{sect2} the basic equations of both pictures
are reviewed. In Sect. \ref{sect3} a 
particular mechansim for dissipation is presented and the
dissipative quantum field theory
calculation is done to determine dissipative effects in the
inflaton effective equation of motion.
In Sect. \ref{sect4} density fluctuations in cold and warm
inflation are presented.  A calculation of inflation for a momomial
potential is done, and it is shown that in contrast to cold
inflation, in the warm inflation case the mass of the inflaton
is bigger than the Hubble parameter and the field amplitude is
below the Planck scale.  These are two very different outcomes
of warm inflation that are not found in the cold inflation case
for such potentials.  In Sect. \ref{susyhy} an inflation
analysis of the SUSY hybrid model is done.  The parameter space
is shown to divide into regimes of cold and warm inflation.
The spectral index is also calculated to examine observational 
differences in the two cases.  Finally some examination is made of
particle physics of this model in the warm inflation case.
Finally in Sect. \ref{sect6} our conclusions are given.

\section{Two inflation dynamics}
\label{sect2}

There are two distinct dynamical realizations of inflation.
In the original picture, termed cold, supercooled or isentropic
inflation
\cite{oldi,ni,ci},
the universe rapidly supercools during inflation and
subsequently a reheating phase is invoked to end inflation and
put the universe back into a radition dominated regime.
In this picture, inflaton dynamics is governed by the equation
\begin{equation}
{\ddot \varphi} + 3H {\dot \varphi} +
\xi {\rm R} \varphi + \frac{dV_{\rm eff}(\varphi)}{d\varphi} = 0 ,
\label{cidyn}
\end{equation}
where
$R = 6({\ddot a}/{a} + {\dot a^2}/{a^2})$
is the curvature scalar.
The conditions for slow-roll inflation in this picture require
\begin{equation}
\epsilon_H =
\frac{m_P^2}{2}\left(\frac{V^\prime}{V}\right)^2 \ll 1
\label{epsh},
\end{equation}
and
\begin{equation}
\eta_H = m_P^2 \frac{V^{\prime \prime}}{V} \ll 1,
\label{etah}
\end{equation}
where $m_P \equiv = m_{pl}/\sqrt{8 \pi}=2.4\times 10^{18}$ GeV
is the reduced Planck mass.
In this slow-roll regime the evolution 
equation (\ref{cidyn}) is well approximated
by
\begin{equation}
3H {\dot \varphi} +
\xi {\rm R} \varphi + \frac{dV_{\rm eff}(\varphi)}{d\varphi} = 0 .
\label{ciappdyn}
\end{equation}

In the other picture, termed warm or nonisentropic inflation
\cite{wi},
dissipative effects are important during the inflation period,
so that radition production occurs concurrently with inflationary
expansion.
Phenomenologically, the inflaton evolution in simple warm inflation models
has the form,
\begin{equation}
{\ddot \varphi} + [3H+\Upsilon(\varphi)] {\dot \varphi} +
\xi {\rm R} \varphi + \frac{dV_{\rm eff}(\varphi)}{d\varphi} = 0 .
\label{amapprox}
\end{equation}
For $\Upsilon =0 $, this equation reduces the the 
familiar inflaton evolution equation
for cold inflation Eq. (\ref{cidyn}), 
but for a nonzero $\Upsilon$, it corresponds to
the case where the inflaton field is dissipating energy
into the universe, thus
creating a radiation component.
The conditions for slow-roll inflation
are modified in the
presence of the extra friction term $\Upsilon$, and we have now:
\begin{eqnarray}
\epsilon_\Upsilon & = & \frac{\epsilon_H}{(1+r)^2} < 1\label{epsups}\,,\\
\eta_\Upsilon & = & \frac{\eta_H}{(1+r)^2} < 1 \label{etaups} \,,
\end{eqnarray}
where $r \equiv \Upsilon/(3 H)$, and $\epsilon_H$, $\eta_H$ are
the slow-roll parameters without dissipation given in
Eqs. (\ref{epsh}) and (\ref{etah}).  In addition, when the friction term
$\Upsilon$ depends on the value of the inflaton field, we can define
a third slow-roll parameter
\begin{equation}
\epsilon_{H \Upsilon} = \frac{r}{(1+r)^3} \beta_\Upsilon  < 1 \label{epsupsi}\,,
\end{equation}
with
\begin{equation}
\beta_\Upsilon = \frac{V^\prime}{3 H^2}\frac{
\Upsilon^\prime}{\Upsilon}\,.
\end{equation}
In this slow-roll regime of warm inflation the infl;aton evolution
equation is well approximated by
\begin{equation}
[3H+\Upsilon(\varphi)] {\dot \varphi} +
\xi {\rm R} \varphi + \frac{dV_{\rm eff}(\varphi)}{d\varphi} = 0 .
\label{wiappdyn}
\end{equation}

The dissipation of the inflaton's motion is associated with the production of
entropy. The entropy density of the radiation $s(\phi,T)$ is defined by
a thermodynamic relation in terms of the thermodynamic potential,
\begin{equation}
s=-V_{,T}.
\end{equation}
The rate of entropy production can be deduced from the conservation of
energy-momentum. The total density $\rho$ and pressure $p$ are given by,
\begin{eqnarray}
\rho&=&{\textstyle\frac12}\dot\phi^2+V+Ts\label{density}\\
p&=&{\textstyle\frac12}\dot\phi^2-V.
\end{eqnarray}
Energy-momentum conservation,
\begin{equation}
\dot\rho+3H(p+\rho)=0,
\label{econs}
\end{equation}
now implies entropy production. Making use of Eq. (\ref{amapprox}) 
(with $\xi=0$) we get
\begin{equation}
T(\dot s+3Hs)=\Upsilon\dot\phi^2.\label{wis}
\end{equation}
The zero curvature Friedman equation completes the set of differential
equations for $\phi$, $T$ and the scale factor $a$,
\begin{equation}
3H^2=8\pi G({\textstyle\frac12}\dot\phi^2+V+Ts).\label{wia}
\end{equation}

The entropy production has been described in a slightly different way in
the initial warm inflation papers \cite{wi,Berera:1999ws}. We can recover an
alternative equation in the case when the temperature corrections to the
potential are negligeable. If we set $\delta m_T=0$ in 
the finite temperature effective potential $V_{eff}(\phi,T)$, then
the
radiation density $\rho_r=4sT/3$, and Eq. (\ref{wis}) becomes
\begin{equation}
\dot \rho_r+4H\rho_r=\Upsilon\dot\phi^2.\label{wir}
\end{equation}
This equation is {\it only} valid when $\delta m_T=0$.

Eq. (\ref{wiappdyn}) is the simplest form of warm inflation dynamics.
The basic idea of warm inflation is that radiation production is
occurring concurrently with inflationary expansion due to dissipation from
the inflaton field system.   This dissipation would imply nonconservative
terms in the inflaton evolution equation.  In gneeral these terms need
not be of the simple temporally local form $\Upsilon {\dot \varphi}$ as in 
Eq. (\ref{wiappdyn}), but could be nonlocal.  One such example would be
a form like 
$\int K(t,t') \varphi^n(t') dt'$.
Moreover in general this radiation production could be
produced far from equilibrium.

\section{Radiation production mechanism during inflation}
\label{sect3}

The key question is what types of first principles mechanisms can result
in radiation production during inflation.  Here we consider the
example of \cite{BR, BR4}, which develops
the mechanism of the scalar inflaton field $\phi$
exciting a heavy bosonic field $\chi$ which then decays to light
fermions $\psi_d$,
                                                                                
\begin{equation}
\phi \rightarrow \chi \rightarrow \psi_d.
\label{wimech}
\end{equation}
This mechanism is expressed in its simplest form
by an interaction Lagrangian density for the coupling of the
inflaton field to the other fields of the form
                                                                                
\begin{equation}
{\cal L}_I =  -\frac{1}{2}g^2 \Phi^2 \chi^2 -
g' \Phi {\bar \psi_{\chi}} \psi_{\chi} - h \chi {\bar \psi_d}\psi_d ,
\label{lint}
\end{equation}
                                                                                
\noindent
where $\psi_d$ are the light fermions to which $\chi$-particles can
decay, with $m_\chi > 2m_{\psi_d}$. Aside from the last term in Eq.
(\ref{lint}), these are the typical interactions commonly used in
studies of reheating after inflation \cite{reheato,reheat}. However a
realistic inflation model often can also have additional interactions
outside the inflaton sector, with the inclusion of the light fermions
$\psi_d$ as depicted above being a viable option;
an example of this is the SUSY hybrid model for which a
recent study of warm inflation has been done \cite{bb}
and will be discussed in Sect. \ref{susyhy}.
Moreover in minimal
supersymmetry (SUSY) extensions of the typical reheating model or
multifield inflation models, the interactions of the form as given in
Eq. (\ref{lint}) can emerge as an automatic consequence of the
supersymmetric structure of the model. Since in the moderate to strong
perturbative regime, reheating and multifield inflation models will
require SUSY for controlling radiative corrections, Eq. (\ref{lint})
with inclusion of the $\psi_d$ field thus is a toy model representative
of many realistic situations.

The effective evolution equation for the inflaton background field
arising from mechanism Eq. (\ref{lint}) has been computed 
in \cite{BR,BR4}.
The basic idea is
we are interested in obtaining the effective equation of motion (EOM)
for a scalar field configuration $\varphi = \langle \Phi \rangle$ after
integrating out the $\Phi$ fluctuations, the scalars $\chi_j$ and
spinors $\psi_k,\bar{\psi}_k$. This is a typical ''system-environment''
decomposition of the problem in which $\varphi$ is regarded as the
system field and everything else is the environment, which in particular
means the $\Phi$ fluctuation modes, the scalars $\chi_j$ and the spinors
$\psi_k,\bar{\psi}_k$ are regarded as the environment bath. In a
Minkowski background, at $T=0$, the EOM for $\varphi$ has been derived
in \cite{BR} using the Schwinger closed time path formalism. Here we
follow a completely analogous approach and derive the EOM in a FRW
background following \cite{BR4}. 
The field equation for $\Phi$ can be readily obtained from
Eq. (\ref{lint}) and it is given by
                                                                                
\begin{eqnarray}
& & \ddot{\Phi} + 3 \frac{\dot{a}}{a} \dot{\Phi}
-\frac{\nabla^2}{a^2} \Phi + m_\phi^2 \Phi + \frac{\lambda}{6} \Phi^3
+\xi R \Phi
\nonumber \\
& &  + \sum_{j=1}^{N_{\chi}} g_j^2 \Phi (x) \chi_j^2(x)  = 0 \;.
\label{eqphi1}
\end{eqnarray}
In order to obtain the effective EOM for $\varphi$, we use the
tadpole method.  In this
method we split $\Phi$ in Eq.
(\ref{eqphi1}), as usual, into the (homogeneous) classical expectation
value $\varphi(t) =\langle \Phi \rangle$ and a quantum fluctuation
$\phi(x)$, $\Phi(x) = \varphi(t) + \phi(x)$. In this
case, the field equation for
$\Phi$, after taking the average (with $\langle \phi (x) \rangle=0$),
becomes
                                                                                
\begin{eqnarray}
\ddot{\varphi}(t) & + & 3 \frac{\dot{a}(t)}{a(t)} \dot{\varphi}(t)+
m_\phi^2 \varphi(t) +
\frac{\lambda}{6} \varphi^3(t) 
\nonumber \\
& + & \xi R(t) \varphi(t)
+\frac{\lambda}{2} \varphi(t) \langle \phi^2 \rangle
+\frac{\lambda}{6} \langle \phi^3 \rangle \nonumber \\
& + & \sum_{j=1}^{N_{\chi}} g_j^2 \left[\varphi (t) \langle \chi_j^2 \rangle +
\langle \phi \chi_j^2 \rangle \right] = 0 \;,
\label{eqphi2}
\end{eqnarray}

\noindent
where $\langle \phi^2 \rangle$, $\langle \phi^3 \rangle$, $\langle
\chi_j^2 \rangle$ and $\langle \phi \chi_j^2 \rangle$ can be
expressed \cite{BR} in terms
of the coincidence limit of the (causal) two-point Green's functions
$G^{++}_\phi (x,x')$ and $G^{++}_{\chi_j} (x,x')$, for the $\Phi$ and
$\chi_j$ fields respectively. 
In particular
\begin{eqnarray}
&& G_{\chi_j}^{++}(x,x') = i \langle T_+ \chi_j(x) \chi_j(x') \rangle
\nonumber \\
&& G_{\chi_j}^{--}(x,x') = i \langle T_- \chi_j(x) \chi_j(x') \rangle
\nonumber \\
&& G_{\chi_j}^{-+}(x,x') = i \langle \chi_j(x) \chi_j(x') \rangle
\nonumber \\
&& G_{\chi_j}^{+-}(x,x') = i \langle \chi_j(x') \chi_j(x) \rangle,
\label{Gofchi}
\end{eqnarray}
and similarly for $G_{\phi}(x,x')$.
The momentum space Fourier transfor of $G_{\chi_j}(x,x')$ is 
\begin{eqnarray}
G_{\chi_j}(x,x') & = & i  \int \frac{d^3 q}{(2 \pi)^3}
e^{i {\bf q} . ({\bf x} - {\bf x}')}
\nonumber \\
& & \left(
\begin{array}{ll}
G^{++}_{\chi_j}({\bf q}, t,t') & \:\: G^{+-}_{\chi_j}({\bf q}, t,t') \\
G^{-+}_{\chi_j}({\bf q}, t,t') & \:\: G^{--}_{\chi_j}({\bf q}, t,t')
\end{array}
\right) \: ,
\label{Gmatrix}
\end{eqnarray}
\noindent
where
\begin{eqnarray}
G^{++}_{\chi_j}({\bf q} , t,t') & = & G^{>}_{\chi_j}({\bf q},t,t')
\theta(t-t') 
\nonumber \\
& + & G^{<}_{\chi_j}({\bf q},t,t') \theta(t'-t) ,
\nonumber \\
G^{--}_{\chi_j}({\bf q} , t,t') & = & G^{>}_{\chi_j}({\bf q},t,t')
\theta(t'-t) 
\nonumber \\
& + & G^{<}_{\chi_j}({\bf q},t,t') \theta(t-t') ,
\nonumber \\
G^{+-}_{\chi_j}({\bf q} , t,t') & = & G^{<}_{\chi_j}({\bf q},t,t') ,
\nonumber \\
G^{-+}_{\chi_j}({\bf q},t,t') & = & G^{>}_{\chi_j}({\bf q},t,t')\;
\label{G of k}
\end{eqnarray}
and similarily for $G_{\phi}(x,x')$.
The Green's functions  $G^{>,<}_{\chi_j}({\bf q},t,t')$ are
written in terms of the modes of the scalar field as
\begin{eqnarray}
G_{\chi_j}^{>} ({\bf q},t,t') & = &
f_{\chi_j;1} ({\bf q},t) f_{\chi_j;2}({\bf q},t') \theta(t-t')
\nonumber \\
& + & f^{*}_{\chi_j;1} ({\bf q},t') f^{*}_{\chi_j;2}({\bf q},t)  \theta(t'-t)\;,
\nonumber \\
G_{\chi_j}^{<} ({\bf q},t,t') & = &
f_{\chi_j;1}^* ({\bf q},t) f_{\chi_j;2}^*({\bf q},t') \theta(t-t')
\nonumber \\
& + & f_{\chi_j;1} ({\bf q},t') f_{\chi_j;2}({\bf q},t)  \theta(t'-t)\;,
\label{prop-terms}
\end{eqnarray}
where in general there will be two independent 
solutions $f_{\chi_j;1,2}({\bf q},t)$ since
$\chi_j$ obeys an equation second order in time.
To obtain the evolution equations for the mode 
functions in Eq. \ref{prop-terms},
the fermion fields that interact with $\chi_j$ fields 
are integrated out following \cite{BR4}, which then leads to
\begin{eqnarray}
& & \left[\frac{d^2}{dt^2} + 3 \frac{\dot{a}}{a}
\frac{\partial}{\partial t} + \frac{\bf{q}^2}{a^2} + M_{\chi_j}^2 (t) \right]
f_{\chi_j}({\bf q},t) 
\nonumber \\
& & +
\int d t' a^3(t') \Pi_{\chi_j} ({\bf q};t,t')  f_{\chi_j}({\bf q},t') =0 \;,
\label{modes chi}
\end{eqnarray}
where $\Pi_{\chi_j}  ({\bf q};t,t')$ is the spatial
{}Fourier transform of the $\chi_j$
field self-energy term.  In particular, in terms of the self-energy
matrix on the closed time path $\Pi_{\chi_j}$ is defined through
\begin{eqnarray}
&& \Sigma_{\chi_j}^{++} (x,x') + \Sigma_{\chi_j}^{+-} (x,x') -
\Sigma_{\chi_j}^{-+} (x,x') - \Sigma_{\chi_j}^{--} (x,x')
\nonumber \\
&& = 2 \left[ \Sigma_{\chi_j}^{++} (x,x') + \Sigma_{\chi_j}^{+-} (x,x')
\right] 
\nonumber \\
&&= 2 \theta(t_1-t_2) \left[ \Sigma_{\chi_j}^{>} (x,x') -
\Sigma_{\chi_j}^{<} (x,x')
\right] \nonumber \\
&&= \Pi_{\chi_j} (x,x') =\Pi_{1,\chi_j}(x,x') + \Pi_{2,\chi_j} (x,x')\;,
\label{2nd}
\end{eqnarray}

\noindent
where
                                                                                
\begin{eqnarray}
\Pi_{1,\chi_j}(x,x') & = & \left[2 \theta(t_1-t_2) -1\right]
\nonumber \\
& & \left[  \Sigma_{\chi_j}^{>} (x,x') -
\Sigma_{\chi_j}^{<} (x,x')
\right] \;,
\nonumber \\
\Pi_{2,\chi_j}(x,x') & = & \Sigma_{\chi_j}^{>} (x,x') -
\Sigma_{\chi_j}^{<} (x,x') \;,
\label{Pis}
\end{eqnarray}
                                                                                
\noindent
which have the properties $\Pi_{1,\chi_j} (x,x') = \Pi_{1,\chi_j} (x',x)$
and $\Pi_{2,\chi_j} (x,x') = -\Pi_{2,\chi_j} (x',x)$.

Typically,  equations for the mode functions for an interacting model,
of the general form as given by Eq. (\ref{modes chi}), can be very
difficult to solve analytically, in particular for an
expanding background. There are a few particular cases,
such as for de Sitter
expansion $H\sim$ constant, so $a(t) = \exp(H t)$, and power law
expansion $a(t) \sim t^n$, where solutions for the mode equation
for free fluctuations are known in exact analytical form
(see e.g. Ref. \cite{davis}).
However, for deriving an approximate solution for the mode
functions in the interacting case, we can apply a WKB
approximation for equations of the general
form Eq. (\ref{modes chi}) and then check the validity of the
approximation for the parameter and dynamical regime
of interest to us.
As will be seen below, under
the dynamical conditions we are interested in studying in this paper,
this approximation will suit our purposes. Let
us briefly recall
the WKB approximation and its general validity regime, when applied to
obtaining approximate solutions for field mode equations.
An approximated WKB solution for a mode equation like
                                                                                
\begin{eqnarray}
\left[ \frac{d^2}{dt^2} + \omega^2 ({\bf q},t) \right]
f({\bf q}, t)  =0 \;,
\label{f wkb}
\end{eqnarray}
                                                                                
\noindent
is of the form
\begin{equation}
f_{WKB}({\bf q},t)= 1/\left[\omega({\bf q},t)\right]^{1/2}
\exp\left[ \pm i \int^t dt'' \omega({\bf q},t'')\right],
\end{equation}
which holds under the general
adiabatic condition $\dot{\omega}({\bf q},t) \ll \omega^2({\bf q},t)$.

Proceeding with our derivation, consider then a differential
equation in the form of Eq. (\ref{modes chi}).
Instead of working in cosmic time, it is more convenient to work
in conformal time $\tau$, defined by $d \tau = dt/a(t)$, in which case
the metric becomes conformally flat,
                                                                                
\begin{equation}
ds^2 = a(\tau)^2 \left( d\tau^2 - d {\bf x}^2 \right)\;,
\end{equation}
By also defining a rescaled mode field in conformal time
by
                                                                                
\begin{equation}
\frac{1}{a(\tau)} \bar{f} ({\bf q}, \tau) = f ({\bf q},t) \;,
\label{mode tau}
\end{equation}

\noindent
we can then re-express Eq. (\ref{modes chi}) in the form (generically
valid for either $\phi$ or $\chi_j$ scalar fluctuations)
                                                                                
\begin{eqnarray}
& & \frac{d^2}{d \tau^2} \bar{f} ({\bf q}, \tau) +
\bar{\omega}({\bf q},\tau)^2 \bar{f} ({\bf q}, \tau)
\nonumber \\
& & + \int d \tau' \bar{\Pi} ({\bf q},\tau,\tau')  \bar{f}({\bf q},\tau') =0 \;,
\label{modes tau}
\end{eqnarray}
                                                                                
\noindent
where we have defined
                                                                                
\begin{equation}
\bar{\omega}({\bf q},\tau)^2 = {\bf q}^2+ a(\tau)^2 \left[M^2 +
\left(\xi - \frac{1}{6} \right) R (\tau) \right] \; .
\label{omegabar}
\end{equation}
In (\ref{omegabar}) the conformal symmetry appears in
an explicit form, with $\xi =1/6$ referring to fields conformally coupled to
the curvature, while
$\xi=0$ gives the minimally coupled case. Note also that in conformal time
the scalar curvature becomes
                                                                                
\begin{equation}
R(\tau) = \frac{6}{a^3} \frac{d^2 a}{d\tau^2}\;.
\end{equation}
                                                                                
\noindent
In Eq. (\ref{modes tau}) we have also
defined the self-energy in conformal time as,
                                                                                
\begin{equation}
\frac{\bar{\Pi} ({\bf q},\tau,\tau')}{a(\tau)^{3/2} a(\tau')^{3/2} } =
\Pi({\bf q},t,t') \;,
\end{equation}

\noindent
where the self-energy contribution  $\Pi$, coming from the integration over
the bath fields, is given by the space {}Fourier transformed form
for Eq. (\ref{2nd}).  In (\ref{2nd}), $\Pi$ was split into
symmetric and antisymmetric pieces with respect to its argument
as defined in Eq. (\ref{Pis}).
Thus based on Eq. (\ref{Pis}), the self-energy term in (\ref{modes tau})
can then be written as
$\bar{\Pi} ({\bf q},\tau,\tau')=\bar{\Pi}_1 ({\bf q},\tau,\tau') +
\bar{\Pi}_2 ({\bf q},\tau,\tau')$.
In addition, by writing the self-energy
term in a diagonal (local) form \cite{ringwald,ian3}
                                                                                
\begin{eqnarray}
\bar{\Pi} ({\bf q},\tau,\tau') & = & \bar{\Pi} ({\bf q},\tau)
\delta(\tau-\tau') 
\nonumber \\
& = & \left[\bar{\Pi}_1 ({\bf q},\tau) +
\bar{\Pi}_2 ({\bf q},\tau)\right]\delta(\tau-\tau')\;,
\end{eqnarray}

\noindent
and from the properties satisfied by $\Pi_1$ and $\Pi_2$, it results
that $\bar{\Pi}_1 ({\bf q},\tau)$ must be real, while
$\bar{\Pi}_2 ({\bf q},\tau)$ must be purely imaginary.
The real
part of the self-energy contributes to both mass and wave function
renormalization terms that can be taken into account by a proper redefinition
of both the field and mass $M$. On the other hand,
the imaginary term of the self-energy is
associated with decaying processes, as discussed previously.
So, we can now  relate  the decay width in terms of the CTP
self-energy terms as

\begin{equation}
\bar{\Gamma}  =
-\frac{{\rm Im} \bar{\Pi}}
{2 \bar{\omega}} = \frac{ \bar{\Sigma}^> - \bar{\Sigma}^< }
{2 \bar{\omega}}\;,
\label{Gammabar}
\end{equation}

\noindent
and Eq. (\ref{modes tau}) can be put in the form

\begin{eqnarray}
\left[ \frac{d^2}{d \tau^2}  +
\bar{\omega}({\bf q},\tau)^2
-  2 i  \bar{\omega}({\bf q},\tau) \bar{\Gamma}({\bf q},\tau) \right]
\bar{f}({\bf q},\tau) =0 \;.
\label{modesgamma}
\end{eqnarray}
We now proceed to obtain a standard WKB solution
for Eq. (\ref{modesgamma}).  To do this, following the usual WKB
procedure, we assume the solution to have the form $\bar{f}({\bf q},\tau) =
c \exp[i \gamma({\bf q}, \tau)]$, where $c$ is some constant that can be fixed
by the initial conditions, given by (\ref{mode_cond}) below.
This form of the solution is then substituted into (\ref{modesgamma})
to give

\begin{equation}
i \gamma'' - \gamma'^2 + \bar{\omega}^2 -2 i \bar{\omega} \bar{\Gamma} =0\;.
\label{eq gamma}
\end{equation}
Working in the standard WKB approximation, for the
zeroth order approximation we
neglect the second derivative term in (\ref{eq gamma}). Then, by taking
$\bar{\Gamma} \ll \bar{\omega}$,
we obtain
                                                                                
\begin{equation}
\gamma_0 \approx \mp \int^\tau_{\tau_0} d\tau' \left(\bar{\omega}-i\bar{\Gamma}\right) \;,
\end{equation}
which is then used in Eq. (\ref{eq gamma}) for the second derivative term to
determine the next order approximation,

\begin{equation}
\gamma_1 \approx \mp \int^\tau_{\tau_0} d\tau' \left[\bar{\omega}-i\bar{\Gamma}
+ {\cal O} \left( \bar{\omega}'^2/\bar{\omega}^3 \right) \right]
+i\ln \sqrt{\bar{\omega}} \;.
\end{equation}
The next and following orders in the approximation bring higher
powers and derivatives of $\bar{\omega}'/\bar{\omega}^2 $, which
in the adiabatic regime, $\bar{\omega}'/\bar{\omega}^2 \ll 1$, are
negligible and we are then led to the result
                                                                                
\begin{equation}
\bar{f}_{1,2} ({\bf q},\tau) \approx \frac{c}{\sqrt{\bar{\omega}}} \exp\left[\mp i
\int^\tau_{\tau_0} d\tau' \left(\bar{\omega}-i\bar{\Gamma}\right) \right]\;.
\label{F mode}
\end{equation}

The solutions for the modes of the form Eq. (\ref{F mode}) and their complex
conjugate are general
within the adiabatic, or WKB, approximation regime of dynamics.
{}Finally, we completely and uniquely determine the modes by fixing the
initial conditions at
some initial reference time $\tau_0$, which can be chosen such that
in the limit of $k
\to \infty$ or $H \to 0$ we
reproduce the Minkowski results. These conditions, which correspond to
the ones for the Bunch-Davis vacuum \cite{davis},
can be written as

\begin{eqnarray}
&& \bar{f}_{1,2}({\bf q},\tau_0)= \frac{1}{\sqrt{2\bar{\omega}(\tau_0)}} \;,
\nonumber \\
&& \bar{f\;}'_{1,2}({\bf q},\tau_0)= \mp i \sqrt{\bar{\omega}(\tau_0)/2}\;,
\label{mode_cond}
\end{eqnarray}
which already fixes the constant $c$ in (\ref{F mode}) as
$c=1/\sqrt{2}$.
                                                                                
Using the above results in (\ref{prop-terms}) and
after returning to cosmic time $t$, we obtain the result, valid
within the WKB approximation, or adiabatic regime,

\begin{equation}
G^{>(<)} ({\bf q} ,t,t') =  \frac{1}{\left[a(t) a(t')\right]^{3/2} }
\tilde{G}^{>(<)} ({\bf q} ,t,t')\;,
\end{equation}
where

\begin{eqnarray}
\tilde{G}^{>}({\bf q} ,t,t') & = &
\frac{1}{2 [\omega(t)\omega(t')]^{1/2}}
\nonumber \\
& & \left\{
e^{-i\int_{t'}^t dt'' [\omega(t'')-i\Gamma(t'')]}
\theta(t-t') \right. 
\nonumber \\
& + & \left.
e^{-i\int_{t'}^t dt'' [\omega(t'')+i\Gamma(t'')]}
\theta(t'-t) \right\}\;,
\nonumber \\
\tilde{G}^{<}({\bf q} , t,t') & = & \tilde{G}^{>}({\bf q}, t',t) \: ,
\label{G><}
\end{eqnarray}

\noindent
where $\Gamma$ is the field decay width in cosmic time,
obtained from (\ref{Gammabar}) and
                                                                                
\begin{equation}
\omega(t) = \sqrt{\frac{{\bf q}^2}{a(t)^2} + M^2(t)}\;,
\end{equation}
with $M^2(t)$, for $\Phi$ particles, given by
                                                                                
\begin{equation}
M^2_\phi(t) = m_\phi^2 +  \frac{\lambda}{2} \varphi(t)^2 + \left(\xi-
\frac{1}{6} \right) R(t)\;,
\label{Mphi}
\end{equation}
while for $\chi_j$ particles,
                                                                                
\begin{equation}
M^2_{\chi_j}(t) = m_{\chi_j}^2 +  g_j^2 \varphi(t)^2 + \left(\xi-
\frac{1}{6} \right) R(t)\;.
\label{Mchi}
\end{equation}

The same result Eq. (\ref{G><}) could in principle be inferred in an
alternative way by expressing the propagator expressions in terms of a
spectral function, defined by a {}Fourier transform for the difference
between the retarded and advanced dressed propagators, 
\begin{eqnarray}
G^{\rm ret}(x,x') & = & \theta(t-t') \left[
G^>(x,x') - G^{<}(x,x') \right]
\nonumber \\
& = &
G^{++}(x,x') - G^{+-}(x,x')\;,
\label{ret}
\\
G^{\rm adv}(x,x') & = & \theta(t'-t) \left[
G^<(x,x') - G^{>}(x,x') \right]
\nonumber \\
& = & G^{++}(x,x') - G^{-+}(x,x')\; ,
\label{adv}
\end{eqnarray}
and approximating the spectral function as
a standard Breit-Wigner form with width given by $\Gamma$ and poles
determining the arguments of the exponential in (\ref{F mode}) and its
complex conjugate \cite{GR}. The validity of this approximation in
particular was recently numerically tested and verified in Ref.
\cite{berges1} for a $1+1\;d$ scalar field in Minkowski space-time. In
the Minkowski space-time case, results analogous to Eq. (\ref{G><}) were
explicitly derived in Refs. \cite{GR,ian,ian1,BGR,BR}. Indeed, for the case
of no expansion $a(t) = {\rm constant}$, Eq. (\ref{G><}) reproduce the
same expressions as found in the case of Minkowski space-time.

The result Eq. (\ref{G><}), from the previous approximations used to derive
the WKB solution Eq. (\ref{F mode}), is valid under the requirements
                                                                                
\begin{eqnarray}
&&\Gamma_\phi \ll \omega_\phi\;,
\nonumber \\
&& \Gamma_{\chi_j} \ll \omega_{\chi_j}\;,
\label{Gammaomega}
\end{eqnarray}
and the adiabatic conditions,
                                                                                
\begin{eqnarray}
&& \frac{\bar{\omega}_\phi'}{\bar{\omega_\phi^2}} =
\frac{\dot{a}/a}{\omega_\phi} + \frac{\dot{\omega}_\phi}{\omega_\phi^2}\ll 1 \;,\nonumber \\
&&
\frac{\bar{\omega}_{\chi_j}'}{\bar{\omega_{\chi_j}^2}} =
\frac{\dot{a}/a}{\omega_{\chi_j}} + \frac{\dot{\omega}_{\chi_j}}{\omega_{\chi_j}^2}\ll 1 \;,
\label{wkb cond}
\end{eqnarray}
                                                                                
\noindent
where in the second term in the equations (\ref{wkb cond}) we have made the change
back to comoving time and used $\bar{\omega} = a(t) \omega (t)= a
\sqrt{q^2/a^2+M^2}$.

We now turn our attention to the EOM Eq. (\ref{eqphi2}), where we will
work it out in the response theory approximation similar to the
treatment in  \cite{BR}. Consider the Lagrangian
density in terms of the background (system) field
$\varphi(t)$ and the fluctuation (bath) fields,
                                                                                
\begin{eqnarray}
{\cal L} [ \Phi & = & \varphi(t)+\phi(x), \chi_j, \bar{\psi}_k, \psi_k, g_{\mu \nu}]
= {\cal L}_\varphi [ \varphi(t), g_{\mu \nu}]
\nonumber \\
& + &
{\cal L}_{\rm bath} [\varphi(t),\phi(x), \chi_j, \bar{\psi}_k, \psi_k, g_{\mu \nu}]\;,
\end{eqnarray}
                                                                                
\noindent
where
                                                                                
\begin{eqnarray*}
& & {\cal L}_\varphi [ \varphi(t), g_{\mu \nu}] =
\nonumber \\
& & a(t)^3 \left\{ \frac{1}{2} \dot{\varphi}(t)^2
-  \frac{m_\phi^2}{2}\varphi(t)^2 -
\frac{\lambda}{4 !} \varphi(t)^4  -\frac{\xi}{2} R \varphi(t)^2
\right\}\;,
\end{eqnarray*}

\noindent
is the sector of the Lagrangian independent of the fluctuation bath
fields, while ${\cal L}_{\rm bath}$ denotes the sector of the Lagrangian
that depends on the bath fields and in particular includes
the key interaction terms Eq. (\ref{lint}). 
In the following derivation it will be
assumed that the background field $\varphi(t)$ is slowly varying,
something that must be checked for self-consistency. Thus, if we
consider the decomposition of $\varphi(t)$ around some arbitrary time
$t_0$ as $\varphi(t) = \varphi(t_0) + \delta \varphi(t)$, $\delta
\varphi(t)$ can be regarded as a perturbation, for which
a response theory approximation can be used for the derivation of the
field averages in Eq. (\ref{eqphi2}).

In response theory we express the change in the expectation value of
some operator $\hat{\cal O}(t)$,
$\delta \langle \hat{\cal O}(t) \rangle =
\langle \hat{\cal O}(t) \rangle_{\rm pert} - \langle \hat{\cal O}(t) \rangle$,
under the influence of some external
perturbation described by $\hat{H}_{\rm pert}$ which is turned
on at some time $t_0$, as (for an introductory
account of response theory, see for instance
Ref. \cite{fetter})
                                                                                
\begin{eqnarray}
\delta \langle \hat{\cal O}(t) \rangle
= i \int_{t_0}^t dt'
\langle \left[ \hat{H}_{\rm pert}(t'), \hat{\cal O}(t)\right]
\rangle_{0}\;,
\label{response0}
\end{eqnarray}
where the expectation value on the RHS of Eq. (\ref{response0}) is evaluated
in the unperturbed ensemble.
The response function defined by Eq. (\ref{response0}) can be readily
generalized for the derivation of the field averages.
Provided that the amplitude $\delta\varphi (t)$ is small
relative to the background field $\varphi(t_0)$,
perturbation theory through the response function can be used to
deduce the expectation values
of the fields that enter in the EOM Eq. (\ref{eqphi2}). In this case
the perturbing Hamiltonian $\hat{H}_{\rm pert}$ is obtained from
${\cal L}_{\rm int}^{\delta \varphi}$, where 
${\cal L}_{\rm int}^{\delta \varphi}$ is the part of the
interaction Largangian proportional to $\delta \varphi$.
{}From ${\cal L}_{\rm int}^{\delta \varphi}$ and Eq. (\ref{response0})
we can then determine the averages of the bath
fields, for example
$\langle \chi_j^2(t)\rangle$, as an expansion in $\delta\varphi(t)$, starting
from the time $t_0$ and in an one-loop approximation, as
\begin{eqnarray}
\langle\chi_j^2 \rangle & \simeq & \langle \chi_j^2 \rangle_0
\nonumber \\
& + & \frac{1}{a(t)^{3}} \int_{t_0}^t
dt' 2 g_j^2 \left[\varphi (t')^2 - \varphi(t_0)^2 \right]
\nonumber \\
& & \int \frac{d^3 {\bf q}}{(2 \pi)^3}
{\rm Im} \left[ \tilde{G}_{\chi_j}^{++} ({\bf q},t,t')
\Bigr|_{\varphi(t_0)} \right]_{t>t'}^2
\;,
\label{lrchi}
\end{eqnarray}
where
\begin{eqnarray}
& &\lefteqn{\langle [\chi_j^2({\bf x},t),\chi_j^2({\bf x},t')]
\rangle = 2 i \: {\rm Im} \langle T \chi_j^2({\bf x},t)\chi_j^2({\bf x},t')\rangle}
\nonumber \\
& & = \frac{4i}{[a(t)a(t')]^{3}}\int \frac{d^3q}{(2 \pi)^3}
{\rm Im}[\tilde{G}_{\chi_j}^{++}({\bf q}, t,t')]_{t>t'}^2\;,
\label{response1}
\end{eqnarray}
Similarly the expression for the other expectation values
$\langle \phi^2 \rangle$, $\langle \phi^3 \rangle$, and
$\langle \phi \chi_j^2 \rangle$ can be determined.

Substituting these field averages into Eq. (\ref{eqphi2}) we then obtain
the $\varphi$-effective equation of motion
\begin{eqnarray}
& & {\ddot \varphi}(t) + 3H(t) {\dot \varphi}(t) +
\frac{dV_{\rm eff}^r(\varphi(t),R(t))}{d\varphi(t)}
\nonumber \\
& & + \lambda^2 \varphi (t) \int_{t_0}^t dt' \:
\varphi(t') \dot{\varphi}(t') K_\phi (t,t')
\nonumber \\
& & + \sum_{j=1}^{N_\chi} 4 g_j^4 \varphi(t)   \int_{t_0}^t
dt'  \varphi(t') \dot{\varphi}(t') K_{\chi_j} (t,t')
= 0\;,
\label{eom3a}
\end{eqnarray}
where $V_{\rm eff}^r$ stand for the renormalized effective potential,
\begin{eqnarray}
K_{\chi_j}(t,t') & = & \int_{t_0}^{t'} dt'' \int \frac{d^3q}{(2\pi)^3}
\sin \left[2\int_{t''}^t d\tau \omega_{\chi_j,t}(\tau)\right]
\nonumber \\
& \times & \frac{\exp\left[-2
\int_{t''}^t d\tau \Gamma_{\chi_j,t} (\tau)\right]}
{4\omega_{\chi_j,t}(t) \omega_{\chi_j,t}(t'')} ,
\label{kernel}
\end{eqnarray}
and similar expression for $K_{\phi}(t,t')$.

In the case where the motion of $\varphi$ is slow, the above equation can
be further simplified.  In particular, in this case the
adiabatic-Markovian approximation can be applied to the nonlocal
kernels  $K_{\chi_j} (t,t')$ etc...
This converts
Eq. (\ref{eom3a}) into one that is completely local in time, albeit with time
derivative terms. The details of this approximation
for Minkowski space-time can be found in \cite{BR}. Its extension
to an expanding FRW background follows analogous lines.
The Markovian approximation amounts to substituting $t' \rightarrow t$ in
the arguments of the $\varphi$-fields in the kernels in
Eq. (\ref{eom3a}).
The adiabatic approximation then requires self-consistently
that all macroscopic motion is
slow on the scale of microscopic motion, thus
${\dot \varphi}/{\varphi},H < \Gamma_{\chi}$.
Moreover when $H < M_{\chi}$,
the kernel $K_\chi(t,t')$ is well approximated by the
nonexpanding limit $H \rightarrow 0$.
The validity of all these approximations were examined
in \cite{BR4}.
The result of these approximations is that, after
trivially integrating over the momentum integral in the last term
in Eq. (\ref{eom3a}), the effective EOM Eq. (\ref{eom3a})
becomes Eq. (\ref{amapprox}).
By setting the couplings $g_j =g'_j=g, h_{kj}=h\sim g \gg \lambda$,
the mass
$M_{\chi} \simeq g\varphi \gg m_{\psi_k}$ and $\Gamma_{\chi} \simeq N_\psi h^2
M_{\chi}^2/[8\pi \omega_{\chi}]$, it leads to the friction coefficient
$\Upsilon(\varphi)$ in Eq. (\ref{amapprox})
                                                                                
\begin{equation}
\Upsilon(\varphi) = N_\chi \frac{\sqrt{2} g^4 \alpha_\chi \varphi^2 }
{64\pi M_\chi \sqrt{1 + \alpha^2_{\chi}}
\sqrt{\sqrt{1 + \alpha^2_{\chi}}+1}},
\label{upsilon}
\end{equation}
where $\alpha_\chi \equiv N_\psi h^2 /(8 \pi )$.

\begin{figure}
\hfil\scalebox{0.45}{\includegraphics{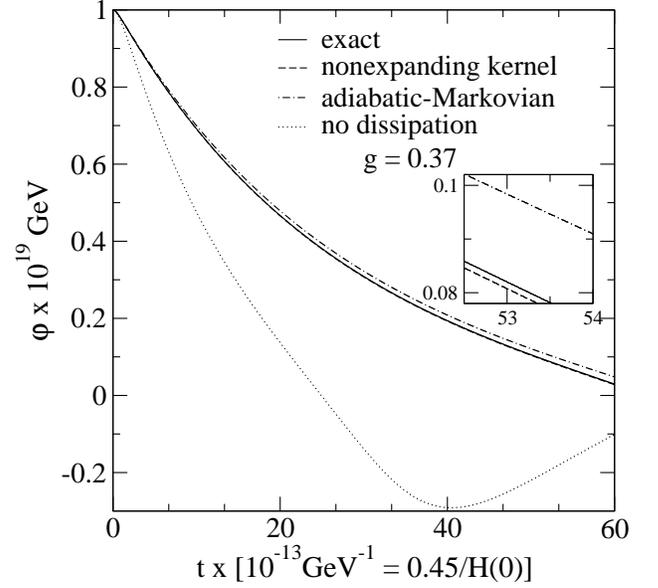}}\hfil
\caption{\label{plot0}
Evolution of $\varphi(t)$ for $\lambda=10^{-13}$,
$g=h=0.37$, $\xi=0$,
$\varphi(0) = m_{\rm Pl}$, ${\dot \varphi}(0)=0$.}
\end{figure}

{}Fig. \ref{plot0} compares the various approximations for a
representative case where $g=h=0.37$ and the inflaton potential is that
for chaotic inflation $V_{\rm eff}(\varphi) = \lambda \varphi^4/4$
with $\lambda = 10^{-13}$ \cite{ci}. In {}Fig \ref{plot0} evolution has
been examined at the final stages of chaotic inflation where we start
with $\varphi(t_0=0) = m_{\rm Pl}$. The solid line is the exact result based on
numerically solving Eq. (\ref{eom3a}). Plotted alongside this, although
almost indiscernible, is the same solution expect using the nonexpanding
spacetime kernel (dashed line), obtained by setting $H \rightarrow 0$,
$a \rightarrow constant$ in Eq. (\ref{kernel}), and the solution based
on the adiabatic-Markovian approximation of Eq. (\ref{amapprox})
(dot-dashed line) for the same parameter set. As seen, the expanding and
nonexpanding cases differ by very little and the adiabatic-Markovian
approximation is in good agreement with the exact solution. This
result presented first in \cite{BR4} confirms simplifying approximations 
claimed in \cite{ringwald,morikawa,BGR,BR,ROR_pascos}
but up to now had not been numerically verified.

More interestingly, the
dotted line in {}Fig. \ref{plot0} is the solution that would be found by
the conventional approach in which the nonlocal terms in Eq. (\ref{eom3a})
are ignored.  The conventional approach \cite{ni,ci,reheato,reheat}
expects the inflaton to start oscillating, which is the precursor to
entering various stages of pre/re-heating. However with account for
dissipative effects, this never happens for our example 
in {}Fig. \ref{plot0},
since the inflaton remains overdamped till the end when it settles at
its minima at $\varphi=0$. Moreover, throughout inflation, and not just
near the end, the inflaton dissipates energy, which yields a
radiation component of magnitude
\begin{equation}
\rho_r \approx \frac{\Upsilon {\dot \varphi}^2}{4H} \;.
\label{rad}
\end{equation}

\section{Density perturbations}
\label{sect4}

The difference between the cold and warm inflationary dynamics
implies that the inflaton density perturbations in the two cases 
also have basic differences.
In particular in the warm inflation regime, since
a thermalized radiation component
is present with $T > m_{\phi}$, inflaton fluctuations are dominantly
thermal rather than quantum.  There are two distinct regimes of
warm inflation to note.  One is the weak dissipative
regime \cite{Moss:wn,Berera:1995wh},
\begin{equation}
\delta \varphi^2  \sim HT \hspace{0.2cm}
{\rm warm \hspace{0.1cm} inflation}
\hspace{0.1cm} (\Upsilon < 3H), \hspace{0.2cm} T > m_{\phi},
\end{equation}
and the other is the strong dissipative regime \cite{Berera:1999ws},
\begin{equation}
\delta \varphi^2 \sim \sqrt{H\Upsilon}T \hspace{0.2cm}
{\rm warm \hspace{0.1cm} inflation} \hspace{0.1cm}(\Upsilon > 3H),
\hspace{0.2cm} T > m_{\phi}.
\label{dpsd}
\end{equation}
For comparison, for cold inflation, where inflaton
fluctuations are exclusively quantum \cite{Guth:ec},
\begin{equation}
\delta \varphi^2 \sim H^2, \hspace{0.2cm}
{\rm cold \hspace{0.1cm}inflation} \hspace{0.2cm} T < m_{\phi} .
\end{equation}
For both cold and warm inflation, density perturbations are obtained by
the same expression,
$\delta \rho/\rho \sim H \delta \varphi/{\dot \varphi}$.

The inherent difference in density perturbations in cold
versus warm inflation provides a possible direction for
distinguishing between these two inflation dynamics using
observation.
In \cite{Taylor:2000ze} an order of magnitude
estimate of density perturbations
during warm inflation was computed by matching the thermally
produced fluctuations to gauge invariant parameters when the fluctuations
cross the horizon (for other phenomenological treatments of warm inflation see
\cite{wipheno}).
This work provided a clear statement of the consistency
condition.
Cold inflation has three parameters,
related to the potential energy
magnitude $V_0$, slope $\epsilon_H$ in Eq. (\ref{epsh})
and $\eta_H$ in Eq. (\ref{etah}), whereas
there are four observable constraints ($\delta_H$,
$A_g$, $n_s$, $n_g$).
This implies a redundancy in the observations and
allows for a consistency relation \cite{Liddle:1993fq}. This is usually
expressed as a relationship
between the tensor-to-scalar ratio and the slope of the tensor spectrum.
Warm inflation has an extra parameter, the dissipation factor,
which implies four constraints for four parameters. Hence we do not expect the
consistency relation of standard inflation
to hold in warm inflation \cite{Taylor:2000ze}.
Thus discriminating between warm and standard
inflation requires measuring
all four observables. The WMAP and upcoming Planck satellite missions
should provide strong constraints on the scalar spectrum
and having polarization detectors, it is hoped
the tensor spectrum also will be measured.
At the same level of approximation,
nongaussian effects from warm inflation models
were computed and found to be of the same
order of magnitude as in the cold inflation case, and
thus too small to be measured \cite{Gupta:2002kn}.

One interesting feauture about warm inflation dynamics is that
it offers a solution to the $\eta$-problem \cite{Berera:2004vm}.
In standard inflation models
\cite{oldi,ni,ci},
where inflaton evolution is damped
by the term $3H {\dot \varphi}$, the slow-roll condition
amounts to $\eta_H \stackrel{<}{\sim} 1$, which equivalently means
the potential can not have mass terms bigger than $\sim H^2 \varphi^2$.

Since Supersymmetry suppresses quantum corrections, thus can
preserve the tree level potential, it has been a central idea
in realizing such flat inflationary potentials.  Of course, since inflation
requires a nonzero vacuum energy density, inevitably
SUSY must be broken during the inflation period,
thus possibly ruining the desired degree of flatness in the potential.
In particular, once supergravity effects are included, it
becomes very difficult for this symmetry to preserve flatness
at the leevl of $\eta_H < 1$.  For F-term inflation, where the nonzero
vacuum energy density arises from terms in the superpotential,
no symmetry prohibits the appearance of the
Planck mass suppressed higher dimensional operators
$a_n \varphi^n/m_{pl}^{n-4}$
\cite{Copeland:1994vg,Gaillard:1995az,Kolda:1998kc,Arkani-Hamed:2003mz}
For the large class of chaotic inflation
type models \cite{ci}, where inflation occurs with the inflaton
field amplitude above $m_{pl}$, to control these higher dimensional
operators would require the fine-tuning of a infinite number
of parameters or choosing only certain types of
SUGRA corrections, such as the mininal Kahler potential.
Even for models where inflation occurs
for field amplitudes below the Planck scale, dimension six operator
terms of the form $V \varphi^2/m^2_{pl} \sim H^2 \varphi^2$ can
emerge and ruin the desired flatness.  Both minimal and nonminimal
Kahler potentials can lead to such terms \cite{Arkani-Hamed:2003mz}.
 
One possible solution to the $\eta$-problem might be D-term inflation.
In such models, the nonzero vacuum energy arises from the
supersymmetrization of the gauge kinetic energy.
However a closer examination
\cite{Kolda:1998kc,Arkani-Hamed:2003mz}
reveals that
attaining the required degree of flatness makes such models
very restrictive.
                                                                                
Up to now, attempts to solve the eta-problem have sought
symmetries that can maintain this desired degree of flatness.
One of the few that has proven successful is called the
Heisenberg symmetry \cite{Gaillard:1995az}, although it is very restrictive.
Another proposal has been a certain shift
symmetry \cite{Arkani-Hamed:2003mz},
which is particularly interesting as it does not require SUSY.
In common, all attempts so far have one foregone conclusion,
that inflaton dynamics is only viable for $\eta_H < 1$.
However, if the inflaton evolution happened to have a damping term
larger than $3H{\dot\varphi}$, then clearly slow-roll
can be satisfied for $\eta_H > 1$. Such a possibility is
precisely what occurs in warm inflationary dynamics.

To see this let us examine the
warm inflation solution
for the simple potential
\begin{equation}
V = \frac{1}{2} m_{\phi}^2 \phi^2.
\label{quadpot}
\end{equation}
In the cold inflation case, such a model requires an initial
inflaton amplitude $\langle \phi \rangle = \varphi > m_{pl}$.
Moreover, SUSY models that realize a potential like this
inevitably lead to an eta-problem based on the reasons
discussed above.
                                                                                
Let us now treat this model in the warm inflation case.
To focus on the essential
points, our calculations here will be purely phenomenological,
although they can be readily derived from a first principles
quantum field theory calculation as done in Sect. \ref{sect3}.
We consider the case
where the dissipative coefficient in Eq. (\ref{amapprox})
is independent of both $\varphi$ and $T$, $\Upsilon = constant$.
The inflaton initially is at a nonzero field
amplitude $\varphi \ne 0$, thus supporting a vacuum energy.

The background cosmology for models with constant $\Upsilon$
and monomial potentials has been solved exactly \cite{wi}.
From this we find $N_e \approx H\Upsilon/m_{\phi}^2$.
The radiation production is determined from the energy conservation
equation (\ref{econs}).
During warm inflation ${\dot \rho}_r \approx 0$
\cite{wi,Berera:1995wh}, so that
Eq. (\ref{econs}) reduces  to Eq. (\ref{rad}).
Identifying $\rho_r \sim T^4$ permits determination of
the temperature during warm inflation.
Finally, once $T$ is determined, Eq. (\ref{dpsd}) allows
determination of density perturbations.

Combining these expressions, for model Eq. (\ref{quadpot}) with
$\Upsilon = const.$ in Eq. (\ref{amapprox}), gives
\begin{equation}
N_e \approx 2\sqrt{2} \frac{\Upsilon \varphi_0}{m_{\phi} m_{pl}}
\end{equation}
\begin{equation}
T \approx \frac{m_{\phi}^{3/4} m_{pl}^{1/4} \varphi_0^{1/4}}{\Upsilon^{1/4}}
\end{equation}
\begin{equation}
\frac{\delta \rho}{\rho} \approx \left(\frac{\varphi_0}{m_{\phi}}\right)^{3/8}
\left(\frac{\Upsilon}{m_{pl}}\right)^{9/8}.
\end{equation}
Imposing observational constraints $N_e =60$ and
$\delta \rho/\rho = 10^{-5}$, leads to the results
\begin{equation}
\frac{m_{\phi}}{H} \approx 5.5 \times 10^{-9} \frac{m_{pl}}{m_{\phi}},
\end{equation}
\begin{equation}
\frac{\varphi_0}{m_{pl}} \approx 5.3 \times 10^{8} \frac{m_{\phi}}{m_{pl}},
\end{equation}
$\Upsilon \approx 4 \times 10^{-8} m_{pl}$,
and $T \approx 10^4 m_{\phi}$, with the ratio $m_{\phi}/m_{pl}$
free to set.  For $m_{\phi}/m_{pl} \stackrel{<}{\sim} 10^{-9}$, it means
$\eta_H > 1$ and $\varphi < m_{pl}$.
Thus we see for sufficiently small
inflaton mass, $m_{\phi} \stackrel{<}{\sim} 10^{10}{\rm GeV}$,
there is no eta-problem, since $m_{\phi}\gg H$
and $\varphi < m_{pl}$.  Since this warm
inflation solution works for $\eta_H \gg 1$, SUSY models
realizing simple monomial potentials like Eq. (\ref{quadpot})
do not require any special symmetrie as is the case
for cold inflation models.
The "eta" and large $\varphi$-amplitude problems simply
correct themselves once interactions already present in the
models are properly treated.

\section{Hybrid SUSY model}
\label{susyhy}

Following the analysis of \cite{bb},
Let us apply the results of the previous two sections
to the SUSY hybrid model
matter field $\Delta$ ${\bar \Delta}$
coupled to it
\begin{equation}
W= \kappa S ( \Phi_1 \Phi_2 - \mu^2) + g \Phi_2 \Delta \bar \Delta \,.
\label{superpot2}
\end{equation}
In this model, the inflaton is identified with
the bosonic part of S, $\phi_S$.
The above is a toy model representing an example of how
the basic hybrid model, first term on the RHS, is embedded within
a more complete particle physics model, in this case through
the $\Delta$ fields.
We will show that in the above model both cold and warm inflation
exist and we will determine the parameter regime for
them.  This will then explicitly verify the conclusions
from the recent papers on dissipation \cite{BR,BR4}, that
showed both types of inflationary dynamics could exist.
In both inflationary regimes, we will calculate the scalar spectral
index $n_S-1$, and its running $dn_S/d\ln k$.
With this information, we will then identify the qualitative
and quantitative differences arising from the warm versus
cold regimes.

In this model the heaviest field with mass $m_+$ can decay into the
massless fermionic partners of $\Delta$ and $\bar  \Delta$, with decay rate:
\begin{equation}
\Gamma_+= \frac{g^2}{16 \pi} m_+ \,.
\end{equation}
We have  $\Gamma_+ \propto \phi_S$, and this can be much larger
than the Hubble rate during inflation:
\begin{equation}
\frac{\Gamma_+}{H}= \sqrt{3}\frac{g^2}{16 \pi}
\left(\frac{m_P}{\mu}\right) (x_N^2+1)^{1/2}\,,
\end{equation}
where $x_N \equiv \phi_S/(\sqrt{2} \mu)$.
Having $\Gamma_+/H>1$, all the way up to the end of inflation, only
requires $g > 0.16$ for $\kappa <0.001$ ( $g > 0.01$ for
$\kappa =0.5$). This allows us to apply the
adiabatic-Markovian limit in the effective EOM for the inflaton
background field, Eq. (\ref{amapprox}),
with the dissipative coefficient given now by:
\begin{equation}
\Upsilon \simeq \frac{\pi^2}{2} \left(\frac{\kappa}{4
  \pi}\right)^3 \left(\frac{g^2}{16 \pi}\right) \frac{x_N^2}{(1 +
  x_N^2)^{1/2}} \mu \label{upsilon2} \,,
\end{equation}
and the ratio to the Hubble rate is given by:
\begin{equation}
\frac{\Upsilon}{3 H} \simeq \frac{\kappa^2}{128 \sqrt{3}\pi}
\left(\frac{g^2}{16\pi}\right)\frac{x_N^2}{(1 +
  x_N^2)^{1/2}} \frac{m_P}{\mu} \,,
\end{equation}
which behaves like  $\Upsilon/(3H) \propto x_N\propto \phi_S$, and
so decreases  during inflation.
That is, the evolution of
the inflaton field may change
from being dominated by the friction term $\Upsilon$ to be dominated
by the Hubble rate $H$. Whether the transition between these two
regimes happens before or after 60 e-folds will depend on the value of
the parameters of the model like $\kappa$ and $g$. The amount of
``radiation'' obtained through the dissipative term,
is given by:
\begin{eqnarray}
\frac{\rho_R}{H^4} &\simeq& \frac{9}{2}\frac{r}{(1+r)^2} \epsilon_H
\frac{m_P^4}{\kappa^2 \mu^4} \label{rhorh4}\,,
\end{eqnarray}
which even when $\Upsilon < H$ could give rise to a thermal bath
with $T > H$. In particular, we can have:
                                                                                
(a) $\Upsilon > 3 H$, and $T > H$  ($\dot \phi_S \simeq
-V_\phi/ \Upsilon$):
\begin{equation}
 \frac{\rho_R}{H^4} \simeq
 \frac{36 \sqrt{3}}{\pi^2}\frac{1}{g^2}
 \left(\frac{m_P}{\mu}\right)^5 \frac{1}{x_N^3}\,,
\end{equation}
                                                                                
(b) $\Upsilon < H$ ($\dot \phi_S \simeq -V_\phi/ (3 H)$):
\begin{equation}
 \frac{\rho_R}{H^4} \simeq
\frac{9}{256 \sqrt{3} \pi}\left(\frac{\kappa}{4 \pi}\right)^4
 \left(\frac{g^2}{16 \pi}\right)
\left(\frac{m_P}{\mu}\right)^7 \frac{1}{x_N}
 \,.
\end{equation}
Note one might expect the presence of this thermal bath may
induce thermal corrections to $\Upsilon$ and $V_{eff}$ but
as shown in \cite{hm}, these corrections are negligible.

\begin{figure}[t]
\hfil\scalebox{0.45} {\includegraphics{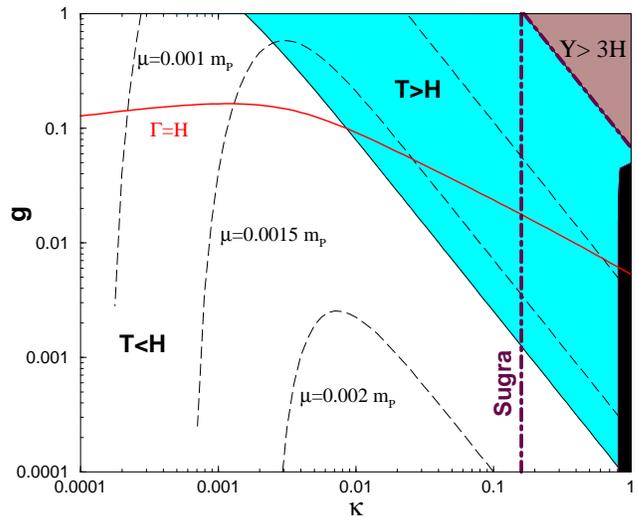}}\hfil
\caption{\label{plot1} 
Regions of cold ($T< H$), and warm ($T > H$) SUSY hybrid
inflation in the $\kappa-g$ plane. The warm inflation region is
divided into the weak dissipative regime with $T > H$ and $\Upsilon_S <
3H$ (lighter shaded region), and the strong dissipative regime with
$\Upsilon_S > 3H$ (darker shaded region). Included are also the
contour plots of constant $\mu$, and the adiabatic-Markovian limit
$\Gamma_+= H$. The black region on the right of the plot is excluded
because $\phi_S > m_P$. In addition, when SUGRA corrections are
taking into account, values to  the left and down
the wide dot-dashed line are excluded.}
\end{figure}

The values of the couplings $\kappa$ and $g$ for which we
could have cold or warm inflation, and strong or weak dissipative
dynamics, are plotted in Fig. (\ref{plot1}). In order to get the
different regions, we have proceeded as follow: for each pair of
values in the plane $\kappa-g$, the value of the inflaton field at the
end of inflation is determined.  This is done in the cold and weak dissipative
regimes by the condition\footnote{The value of $\eta_\Upsilon$ becomes
larger than 1 before the other two slow-roll parameters.}
$\eta_\Upsilon=1$, Eq.(\ref{etaups}). In the strong dissipative regime
inflation can end either with $\eta_\Upsilon =1$ or
it may also happen that most of the vacuum energy is already transferred into
radiation during inflation, and then inflation will end when
$\rho_R \simeq \kappa^2 \mu^4$ instead.
In this case, whichever occurs first fixes the value of
the inflaton field at the end of inflation.
The value of the inflaton field
at 60 e-folds of inflation is then obtained from
\begin{equation}
N_e \simeq - \int_{\phi_{Si}}^{\phi_{Se}} \frac{3 H^2}{ \Delta V^\prime} (
1 + r) d \phi \,. \label{Neups}
\end{equation}
This in turn fixes the value of the dissipative
coefficient $\Upsilon$, Eq. (\ref{upsilon2}), the temperature of thermal
bath, Eq. (\ref{rhorh4}), and therefore the amplitude of the spectrum
$P_{\cal R}$.  The COBE normalization is then used to fix the
value of the scale $\mu$.
In order to match the expressions for the spectrum across
the different regimes, we have used a simple expression with :
\begin{eqnarray}
P^{1/2}_{\cal R} & = &
\left|\frac{3 H^2}{\Delta V^\prime}\right| (1+r) 
\left( 1 +
 \sqrt{\frac{T}{H}} \right) 
\nonumber \\
& & \times \left( 1 +\left(\frac{\pi
   \Upsilon}{4H}\right)^{1/4} \right)\left(\frac{H}{2 \pi}\right) \,.
\end{eqnarray}

We can see in Fig. (\ref{plot1}) that the strong dissipative regime
$\Upsilon > 3H$ requires large values of the couplings, $\kappa \sim g
\sim O(1)$; for values $\kappa \simeq g \simeq 0.1$ we are in the weak
dissipative regime; and for values $\kappa \simeq g \simeq 0.01$ we
recover the cold inflationary scenario.
Typically, for a fixed value of the scale $\mu$ the amplitude of the
spectrum in the strong dissipative regime would be larger than the one
generated at zero $T$. The COBE normalization implies then a smaller
value of the inflationary scale
$\mu$. For example, for $\kappa = g =1$ we have $\mu\simeq 10^{13}$ GeV,
whereas pushing the coupling toward its perturbative limit, $\kappa
=g= \sqrt{4 \pi}$ we would get $\mu \simeq 2 \times 10^{10}$ GeV.
On the other hand, going from the cold to the weak dissipative regime,
the value of $\mu$ only varies by a factor of 2 or 3, and it is still
in the range of the GUT scale $O(10^{15})$ GeV.

\begin{figure}
\hfil\scalebox{0.45}{\includegraphics{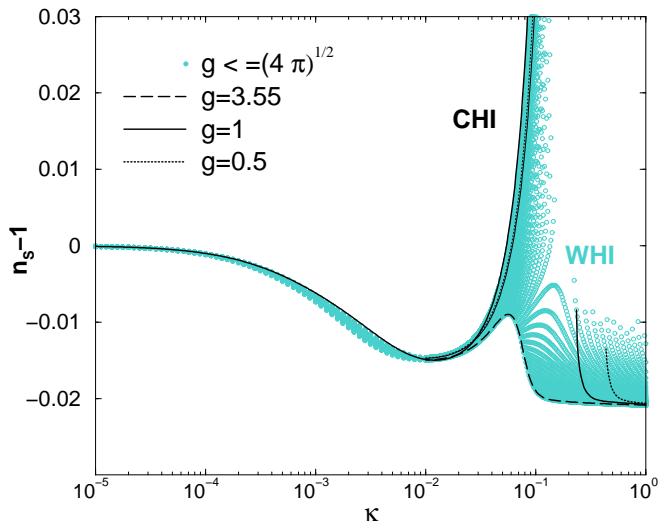}}\hfil
\caption{\label{plot2}
Spectral index for cold SUSY hybrid inflation (solid line, CHI),
and warm inflation (gray region, WHI). The weak dissipative regime ($T >
H$ but $\Upsilon_S < 3H$) is given by the darker gray region
(triangle down);the  strong dissipative regime ($\Upsilon_S > 3H$)
is given by the light gray region
(triangle up). }
\end{figure}

In Fig. (\ref{plot2}) we have compared the prediction for the spectral
index of the scalar spectrum of perturbations in both the CHI
scenario, and warm hybrid inflation (WHI). From the warm inflation
scenario we can always recover the CHI prediction by taking $g \ll 1$. In
standard SUSY GUT hybrid inflation, for small values of the coupling
$\kappa$ the spectrum is practically scale invariant, it reaches a
minimum around $\kappa \simeq 0.01$, and then rises due to SUGRA
corrections up to positive values, which are disfavoured by WMAP
results. But in the weak and the strong
dissipative regime, due to the different origin of the spectrum, we
get that the spectral index is still below 1 even for values of the coupling
$\kappa > 0.01$. Specially in the strong dissipative regime, where the
dynamic is such that the inflaton field is well below the Planck
scale and SUGRA corrections are negligible. In that regime the
departure from scale invariance is within the observational value,
with $n_S -1 \simeq -0.022$.

As a well-motivated example \cite{bb}, which combines inflation with
leptogenesis and light neutrino masses \cite{lazalepto,shafilepto},
the inflaton can decay into right handed (s)neutrino fields $\nu_{Ri}$
($i$= family index). The decay proceed through the
non-renormalizable coupling $\Phi_1 \Phi_1 \nu_{Ri} \nu_{Ri}$,
with decay rate,
\begin{equation}
\Gamma_S = \frac{1}{8 \pi} \left(\frac{M_i}{\mu}\right)^2 m_S \,,
\end{equation}
where $M_i$ is the RH (s)neutrino mass.  In the CHI scenario, with $\mu
\simeq O(10^{15})$ GeV, $\kappa \simeq 10^{-2}$, and $m_S
\simeq 10^{13}$ GeV, the gravitino constraint $T_{RH} \leq 10^{9}$ GeV
translates roughly into $M_i\simeq 10^{-3} \mu \sim O(10^{12})
GeV$. Those values are also
consistent with baryogenesis and light neutrino masses
\cite{shafilepto}. This kind of scenario is also viable in the
warm inflationary regime. Being consistent with the observed baryon
asymmetry and the atmospheric neutrino oscillations
does not directly constraint the value of $\kappa$ but the value of
$m_S \sim 10^{13}$ GeV.
In the warm inflationary regime the
value of the scale $\mu$ required for successful inflation reduces as
we moved into the strong dissipative regime, $m_S$ is  of the
order of $10^{13}$ GeV for $\kappa \simeq O(1)$, and the gravitino
constraint gives now
$M_i\simeq 10^{-3} \mu \sim O(10^{10})$ GeV.
Therefore,
a model of $warm$ inflation and leptogenesis without the need of small
couplings would be viable and compatible with observations, in the
strong dissipative regime.

\section{Conclusion}
\label{sect6}

In this talk we have reviewed the basic ideas of warm inflation
and compared them to the standard or cold inflation picture.
An important point that has been attempted to be converyed is that
interactions in a inflation model Lagrangian can have
significant effects during the inflation period.
This point has led to the key result  
that many typical models of inflation,
which had been assumed to yield just cold inflation dynamics,
in fact have regimes of warm inflation.
The realization of this fact leads to many interesting and
new questions.  From the one direction is model building.
In the warm inflation case, there are some new features
that do not exist for cold inflation.  For one the $\eta$-problem
can be elminated simply by the dissipative dynamics and
does not require any further constaints on the particle phyiscs
model, in particular the Kahler potential.  This means greater
flexibility to the high energy properties of the models.
Another new feature is that for monomial potentials,
observationally consistent warm inflation occurs
for field amplitudes below the Planck scale, in constrast to
cold inflation, where it is well known chaotic inflation
requires $\varphi > m_{pl}$.  This difference implies
from the perspective of effective theories, monomial
potentials are accpetable in the strong
dissipative warm inflation regime.

Another interesting direction is determining tests that could
discrimnate warm versus cold inflation from observational
data.  The nature of the density perturbations in the
two dynamics is different, so one should expect to
find different observational signatures in the two cases.
For one the tensor-to-scalar rations differ for the
two cases.  Also, although we have not explored this direction in
great detail in this talk, more careful analysis of
density perturbations done by evolving the complete set of cosmological
perturbation equations was done in \cite{hmb}.
This analysis showed that
dissipative effects can produce a rich variety
of scalar spectra ranging between red and blue
and can also cause oscillations in the scalar power spectrum.
Further understanding of all these possibilities is needed
before firm criteria can be set of for discriminating
between warm and cold inflation behavior in the CMB data.

This work was funded by a Particle Physics and Astronomy Research Council
(PPARC) Advanced Fellowship


\begin{thebibliography}{99}

\bibitem{oldi} A. H. Guth, Phys. Rev D{\bf 23}, 347 (1981);
K. Sato, Phys. Lett. B{\bf 99}, 66 (1981).
                                                                                
\bibitem{ni} A. Albrecht and P. J. Steinhardt, Phys. Rev. Lett.
{\bf 48}, 1220 (1982); A. Linde, Phys. Lett. B{\bf 108}, 389 (1982).
                                                                                
\bibitem{ci} A. Linde, Phys. Lett. B{\bf 129}, 177 (1983).

\bibitem{star} A. A. Starobinsky, Phys. Lett. B {\bf 91}, 99 (1980).

\bibitem{mc} V. F. Mukhanov and G. V. Chibisov,
JETP Lett. {\bf 33}, 532 (1981).

\bibitem{wi} A. Berera,  Phys. Rev. Lett. {\bf 75},
3218 (1995);
Phys. Rev. D{\bf 54}, 2519 (1996);
Phys.\ Rev.\  D{\bf 55}, 3346 (1997).

\bibitem{Berera:1999ws}
A.~Berera,
Nucl.\ Phys.\ B {\bf 585}, 666 (2000).

\bibitem{BR}A. Berera and R. O. Ramos, Phys.
Rev. D{\bf 63}, 103509 (2001);
A. Berera and R. O. Ramos, Phys. Lett. B{\bf 567}, 294 (2003).
                                                                                
\bibitem{BR4} A. Berera and R. O. Ramos,
Phys. Lett. B{\bf 607},1 (2005);
Phys. Rev. D{\bf 71}, 023513 (2005).

\bibitem{reheato} A. Albrecht, P. J. Steinhardt, M. S. Turner
and F. Wilczek, Phys. Rev. Lett. {\bf 48}, 1437 (1982);
A. D. Dolgov and A. D. Linde, Phys. Lett. B{\bf 116}, 329 (1982);
L. F. Abbott, E. Farhi, and
M. B. Wise, Phys. Lett. B{\bf 117}, 29 (1982).
                                                                                
                                                                                
\bibitem{reheat} L. Kofman, A. Linde and A. A. Starobinsky,
Phys. Rev. D {\bf 56}, 3258 (1997); P. B. Greene and L. Kofman,
Phys. Lett. B{\bf 448}, 6 (1999); F. Finelli and R. Brandenberger,
Phys. Rev. D{\bf 62}, 083502 (2000).

\bibitem{bb} M. Bastero-Gil and A. Berera, 
Phys. Rev. D{\bf 71}, 063515 (2005).

\bibitem{davis} N. D. Birrel and P. C. W. Davis, {\it Quantum Fields in Curved
Space} (Cambridge University Press, Cambridge, England, 1982);
S. A. Fulling, {\it Aspects of Quantum Field Theory in
Curved Space Time} (Cambridge University Press, Cambridge, England, 1989).

\bibitem{ringwald}
A. Ringwald, Ann. Phys. (NY) {\bf 177}, 129 (1987).
                                                                                
\bibitem{ian3}  I. D. Lawrie, Phys. Rev. D{\bf 67}, 045006 (2003).

\bibitem{GR} M. Gleiser and R. O. Ramos, Phys. Rev. D{\bf 50}, 2441
(1994).

\bibitem{berges1}G. Aarts and J. Berges, Phys. Rev. D{\bf 64},
105010 (2001).

\bibitem{ian} I. D. Lawrie, J. Phys. A{\bf 25}, 6493 (1992).

\bibitem{ian1} I. D. Lawrie, Phys. Rev. D{\bf 60}, 063510 (1999).

\bibitem{BGR} A. Berera, M. Gleiser and R. O. Ramos, Phys. Rev. {\bf D58},
123508 (1998);
Phys. Rev. Lett.
{\bf 83}, 264 (1999).
 
\bibitem{fetter}A. L. Fetter and J. D. Walecka, {\it Quantum
Theory of Many Particle Systems} (McGraw-Hill, New York, 1971).
 
\bibitem{morikawa} M. Morikawa and M. Sasaki,
Prog. Theor. Phys. {\bf 72}, 782 (1984).

\bibitem{ROR_pascos} R. O. Ramos, arXiv: hep-ph/0409353.

\bibitem{Moss:wn}
I.~G.~Moss,
Phys.\ Lett.\ B {\bf 154}, 120 (1985).
              
                                                                
\bibitem{Berera:1995wh}
A.~Berera and L.~Z.~Fang,
Phys.\ Rev.\ Lett.\  {\bf 74},  1912 (1995).
 
\bibitem{Guth:ec}
A.~H.~Guth and S.~Y.~Pi,
Phys.\ Rev.\ Lett.\  {\bf 49}, 1110 (1982).

\bibitem{Taylor:2000ze}
A.~N.~Taylor and A.~Berera,
Phys.\ Rev.\ D {\bf 62}, 083517 (2000).
 
\bibitem{wipheno}
H. P. de Oliveira and  R. O. Ramos,
Phys. Rev. {\bf D57}, 741 (1998);
W.~Lee and L.~Z.~Fang,
Phys.\ Rev.\ D {\bf 59}, 083503 (1999);
M.~Bellini,
Nucl.\ Phys.\ B {\bf 563}, 245 (1999);
H.~P.~De Oliveira and S.~E.~Joras,
Phys.\ Rev.\ D {\bf 64}, 063513 (2001);
L.~P.~Chimento, A.~S.~Jakubi, N.~A.~Zuccala and D.~Pavon,
Phys.\ Rev.\ D {\bf 65}, 083510 (2002);
J.~c.~Hwang and H.~Noh,
Class.\ Quant.\ Grav.\  {\bf 19}, 527 (2002);
W.~Lee and L.~Z.~Fang,
Phys. Rev. D {\bf 69}, 023514 (2004);
J. P. Mimosom A. Nunes, and D. Pavon, astro-ph/0410070;
J. M. Silva and J. A. S. Lima, Int. J. Mod. Phys. D {\bf 13},
1315 (2004). 

\bibitem{Liddle:1993fq}
A.~R.~Liddle and D.~H.~Lyth,
Phys.\ Rept.\  {\bf 231}, 1 (1993).

\bibitem{Gupta:2002kn}
S.~Gupta, A.~Berera, A.~F.~Heavens and S.~Matarrese,
Phys.\ Rev.\ D {\bf 66}, 043510 (2002).

\bibitem{Berera:2004vm}
A.~Berera,
arXiv:hep-ph/0401139.

\bibitem{Copeland:1994vg}
E.~J.~Copeland, A.~R.~Liddle, D.~H.~Lyth, E.~D.~Stewart and D.~Wands,
Phys.\ Rev.\ D {\bf 49}, 6410 (1994).
                                                                                
\bibitem{Gaillard:1995az}
M.~K.~Gaillard, H.~Murayama and K.~A.~Olive,
Phys.\ Lett.\ B {\bf 355}, 71 (1995).
                                                                                
\bibitem{Kolda:1998kc}
C.~F.~Kolda and J.~March-Russell,
Phys.\ Rev.\ D {\bf 60}, 023504 (1999).

\bibitem{Arkani-Hamed:2003mz}
N.~Arkani-Hamed, H.~C.~Cheng, P.~Creminelli and L.~Randall,
JCAP {\bf 0307}, 003 (2003).

\bibitem{hm} L. M. H. Hall and I. G. Moss, hep-ph/0408323.
                
\bibitem{lazalepto} G. Lazarides and N. D. Vlachos, Phys. Lett.
B{\bf 459}, 482 (1999).
                                                                               
\bibitem{shafilepto}
V. N. Senoguz and Q. Shafi, Phys. Lett. B{\bf 582}, 6 (2004).

\bibitem{hmb}
L.~M.~H.~Hall, I.~G.~Moss and A.~Berera,
Phys.\ Rev.\ D {\bf 69}, 083525 (2004);
Phys. Lett. B{\bf 589}, 1 (2004).

                                                                               
\end{thebibliography}
\end{document}